\documentclass{llncs}
\usepackage{graphics}
\usepackage{epsfig}

\usepackage{algorithm}
\usepackage{algorithmic}

\date{}

\begin{document}

\pagestyle{empty} \mainmatter

\title{Automatic Selection of Bitmap Join Indexes \\ in Data Warehouses}

\titlerunning{Data warehouse auto-administration:\\ a strategy for automatic bitmap join index selection}
\author{Kamel Aouiche \and Jerome Darmont \and Omar Boussaid \and Fadila Bentayeb}
\institute{ERIC Laboratory -- University of Lyon 2 \\
5, av. Pierre Mend\`{e}s-France\\
F-69676 BRON Cedex -- FRANCE\\
\email{\{kaouiche, jdarmont, boussaid, bentayeb\}@eric.univ-lyon2.fr}}

\maketitle

\begin{abstract}

The queries defined on data warehouses are complex and use several join
operations that induce an expensive computational cost. This cost becomes even
more prohibitive when queries access very large volumes of data. To improve
response time, data warehouse administrators generally use  indexing techniques
such as star join indexes or bitmap join indexes. This task is nevertheless
complex and fastidious. Our solution lies in the field of data warehouse
auto-administration. In this framework, we propose an automatic index selection
strategy. We exploit a data mining technique ; more precisely frequent itemset
mining, in order to determine a set of candidate indexes from a given workload.
Then, we propose several cost models allowing to create an index configuration
composed by the indexes providing the best profit. These models evaluate the
cost of accessing data using bitmap join indexes, and the cost of updating and
storing these indexes.
%\textbf{Keywords:}  Data warehouses, auto-administration, index selection,
%frequent itemsets, cost models, bitmap join indexes.
\end{abstract}

\section{Introduction}

Data warehouses are generally modelled according to a star schema that contains
a central, large fact table, and several dimension tables that
describe the facts~\cite{inm02bui,kim02dat}. The fact table contains the
keys of the dimension tables (foreign keys) and measures. A decision--support
query on this model needs one or more joins between the fact table and the
dimension tables.
%In addition, the data warehouse schema may contain
%hierarchies on dimensions (snowflake schema), which entails additional joins.
These joins induce an expensive computational cost. This cost becomes even more
prohibitive when queries access very large data volumes. It is thus crucial to
reduce it.

Several database techniques have been proposed to improve the computational
cost of joins, such as hash join, merge join and nested loop
join~\cite{mis92joi}. However, these techniques are efficient only when a join
applies on two tables and data volume is relatively small. When the number of
joins is greater than two, they are ordered depending  on the joined tables
(join order problem). Other techniques, used in the data warehouse environment,
exploit join indexes to pre-compute these joins in order to ensure fast data
access. %~\cite{biz01dim,bri97sta,one95mul,val87joi,wu98enc}.
Data warehouse administrators then handle the crucial task of choosing the best
indexes to create (index selection problem). This problem has been studied for
many years in
databases~\cite{agr00aut,cha04ind,fel03nea,fin88phy,fra92ada,kra03gen,val00db2}.
However, it remains largely unresolved in data warehouses. Existing research
studies may be clustered in two families: algorithms that optimize maintenance
cost~\cite{lab97phy} and algorithms that optimize query response
time~\cite{agr01mat,gol02ind,gup97ind}. In both cases, optimization is realized
under the constraint of the storage space. In this paper, we focus on the
second family of solutions, which is relevant in our context because they aim
to optimize query response time.

In addition, with the large scale usage of databases in general and data
warehouses in particular, it is now very important to reduce the database
administration function. The aim of auto-administrative systems is to
administrate and adapt themselves automatically, without loss (or even with a
gain) in performance. In this context, we proposed a method for index selection
in databases based on frequent itemset extraction from a given
workload~\cite{aou03fre}. In this paper, we present the follow-up of this work.
Since all candidate indexes provided by the frequent itemset extraction phase
cannot be built in practice due to system and storage space constraints, we
propose a cost model--based strategy that selects the most advantageous
indexes. Our cost models estimate the data access cost using bitmap join
indexes, and their maintenance and storage cost.

We particularly focus on bitmap join indexes because they are well--adapted to
data warehouses. Bitmap indexes indeed make the execution of several common
operations such as \textbf{\texttt{And}}, \textbf{\texttt{Or}},
\textbf{\texttt{Not}} or \textbf{\texttt{Count}} efficient by having them
operating on bitmaps, in memory, and not on the original data. Furthermore,
joins are pre-computed at index creation time and not at query execution time.
The storage space occupied by bitmaps is also low, especially when the indexed
attribute cardinality is not high~\cite{sar97ind,wu99que}. Such attributes are
frequently used in decision--support query clauses such as
\textbf{\texttt{Where}} and \texttt{\textbf{Group by}}.

The remainder of this paper is organized as follows. We first remind the
principle of our index selection method based on frequent itemset mining
(Section~\ref{Index selection}). Then, we detail our cost models
(Section~\ref{Cost models}) and our index selection strategy
(Section~\ref{Strategy}). To validate our work, we also present some
experiments (Section~\ref{Experiments}). We finally conclude and provide
research perspectives (Section~\ref{Conclusion}).

\section{Index selection method}\label{Index selection}

\begin{figure}[t]
\centering
   {\epsfig{file = 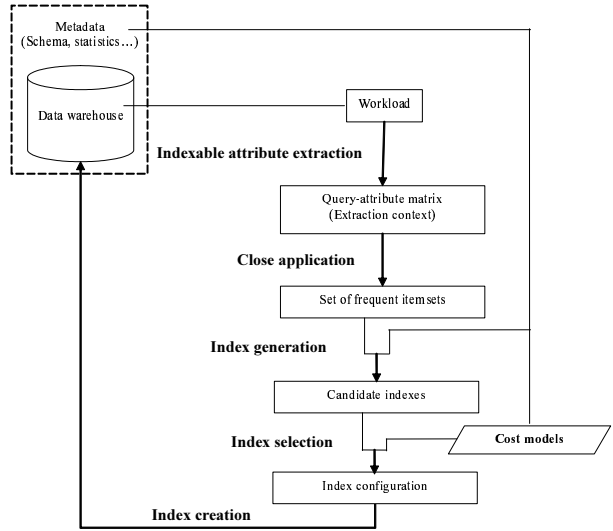, width = \hsize*6/9}}
\caption{Automatic index selection strategy} \label{fig:index_selection}
\end{figure}

In this section, we present an extension to our work about the index selection
problem~\cite{aou03fre}. The method we propose
(Figure~\ref{fig:index_selection}) exploits the transaction log (the set of all
the queries processed by the system) to recommend an index configuration
improving data access time.

We first extract from a given workload a set of so called
indexable attributes. %These attributes are stored in a
%``query-attribute'' matrix, which represents an extraction
%context.
Then, we build a ``query-attribute'' matrix whose rows represent
workload queries and whose columns represent a set of all the
indexable attributes. Attribute presence in a query is symbolized
by one, and absence by zero. It is then exploited by the Close
frequent itemset mining algorithm~\cite{pas99dis}. Each itemset is
analyzed to generate a set of candidate indexes. This is achieved
by exploiting the data warehouse metadata (schema: primary keys,
foreign keys;  statistics\dots). Finally, we prune the candidate
indexes using the cost models presented in Section~\ref{Cost
models}, before effectively building a pertinent index
configuration. We detail these steps in the following sections.

\section{Cost models}\label{Cost models}

The number of candidate indexes is generally as high as the input
workload is large. Thus, it is not feasible to build all the
proposed indexes because of system limits (limited number of
indexes per table) or storage space constraints. To circumvent
these limitations, we propose cost models allowing to conserve
only the most advantageous indexes. These models estimate the
storage space (in bytes) occupied by bitmap join indexes, the data
access cost using these indexes and their maintenance cost
expressed in number of input/output operations (I/Os).
Table~\ref{tab:symbol} summarizes the notations used in our cost
models.

\begin{table}[t]
\caption{Cost model parameters}
\begin{center}
\begin{tabular}{|p{2cm}|p{10cm}|} \hline
\textbf{Symbol} & \textbf{Description} \\ \hline \hline $\mid X\mid$ & Number
of tuples in table X or cardinality of attribute X \\ \hline $S_{p}$ & Disk
page size in bytes\\ \hline $p_{X}$ & Number of pages needed to store table X
\\ \hline $S_{pointer}$ &
Page pointer size in bytes\\ \hline  m & B-tree order \\
\hline d & Number of bitmaps used to evaluate a given query\\
\hline
$w(X)$ & Tuple size in bytes of table X or attribute X \\
\hline
\end{tabular}
\label{tab:symbol}
\end{center}
\end{table}

\subsection{Bitmap join index size}

The space required to store a simple bitmap index linearly depends on the
indexed attribute cardinality and the number of tuples in the table on which
the index is built. The storage space of a bitmap index built on attribute $A$
from table $T$ is equal to $\frac{|A||T|}{8}$ bytes~\cite{wu99que,wu98enc}.
Bitmap join indexes are built on dimension table attributes. Each bitmap
contains as many bits as the number of tuples in fact table $F$. The size of
their storage space is then $S=\frac{|A||F|}{8}$ bytes.

%The construction time of a bitmap join index depends both on the cardinality of
%the indexed attribute and the number of tuples in the indexed table. Complexity
%is then $O(|A||F|)$~\cite{wu99que}.

\subsection{Bitmap join index maintenance cost}

%Update operations have a direct impact on a bitmap join index's size,
%especially when the number of bitmaps is high.

Data updates (mainly insert operations in decisions-support systems)
systemically trigger index updates. These operations are applied either on a
fact table or dimensions. The cost of updating bitmap join indexes is presented
in the following sections.

%The size variation due to update operations is presented in the following
%sections.

%\subsubsection{Bitmap join index size variation.}

%Data updates systemically trigger index updates. This causes storage space
%variations. An update may or may not expand the domain of an indexed attribute.
%There is an expansion when the update adds new values to the attribute domain.
%Otherwise, the update is without expansion. For example, if the domain of
%attribute ``Type'' from table \textbf{\texttt{Products}} is $\{A,B,C\}$, the
%SQL statement \textbf{\texttt{Insert into Products(Type) Values (`K')}} expands
%this attribute's domain.

%\begin{itemize}

%\item \textbf{Update without expansion:}  In this case, a new bit
%corresponding to the inserted tuple is added to each already created bitmap.
%The bit value is set to one in the bitmap corresponding to the inserted tuple
%and to zero in the other bitmaps. The space size variation is $\Delta
%S~=~\frac{|A|(|F|+1)}{8}-\frac{|A||F|}{8}~=~\frac{|A|}{8}$. %and complexity is
%$O(|A|)$~\cite{wu98enc}.

%\item \textbf{Update with expansion:}  The attribute's domain is
%expanded by a new value. A new bitmap corresponding to this value is then
%created. The space variation is $\Delta
%S~=~\frac{(|A|+1)(|F|+1)}{8}-\frac{|A||F|}{8}~=~\frac{|A|}{8}+\frac{|F|+1}{8}$.%and
%complexity is $O(|A|)+O(|F|)$~\cite{wu98enc}.

%\end{itemize}

\noindent \textbf{Insertion cost in fact table. } Assume a bitmap join index
built on attribute $A$ from dimension table $T$. While inserting tuples in fact
table $F$, it is first necessary to search for the tuple of $T$ that is able to
be joined with them. At worst, the whole table $T$ is scanned ($P_{T}$ pages
are read). It is then necessary to update all bitmaps. At worst, all bitmaps
are scanned: $\frac{|A||F|}{8 S_{p}}$ pages are read, where $S_{p}$ denotes the
size of one disk page. The index maintenance cost is then
$C_{maintenance}=p_{T}+\frac{|A||F|}{8 S_{p}}$.

\noindent \textbf{Insertion cost in dimension tables. } An insertion in
dimension $T$ may induce or not a domain expansion for attribute $A$. When not
expanding the domain, the fact table is scanned to search for tuples that are
able to be joined with the new tuple inserted in $T$. This operation requires
to read $p_{F}$ pages. It is then necessary to update the bitmap index. This
requires $\frac{|A||F|}{8 S_{p}}$ I/Os. When expanding the domain, it is
necessary to add the cost of building a new bitmap ($\frac{|F|}{8 S_{p}}$
pages). The maintenance cost of bitmap join indexes is then
$C_{maintenance}~=~p_{F}+(1+\xi)\frac{|A||F|}{8 S_{p}}$, where $\xi$ is equal
to one if there is expansion and zero otherwise.

\subsection{Data access cost}

We propose two cost models to estimate the number of I/Os needed for data
access. In the first model, we do not take any hypothesis about how indexes are
physically implemented. In the second model, we assume that access to the index
bitmaps is achieved through a b-tree such as is the case in Oracle. Due to lack
of space and our experiments under Oracle we only detail here the second model
because of running our experiments. however, the first model is not detailed
here due to the lack of space.

%in Appendix~\ref{model1}.

%\subsubsection{B-tree access to bitmaps}

In this model, we assume that the access to bitmaps is realized through a
b-tree (meta--indexing) in which leaf nodes point to bitmaps
(appendix~figure~\ref{fig:bitmap_join}). The cost, in number of I/Os, of
exploiting a bitmap join index for a given query may be written as follows:
$C~=~C_{descent}~+~C_{scan}~+~C_{read}$, where $C_{descent}$ denotes the cost
needed to reach the leaf nodes from the b-tree root, $C_{scan}$ denotes the
cost of scanning leaf nodes to retrieve the right search key and the cost of
reading the bitmaps associated to this key, and $C_{read}$ finally gives the
cost of reading the indexed table's tuples.

%\begin{figure}[h]
%\centering
 %  {\epsfig{file = bitmap_index.eps, width = \hsize*6/9}}
%\caption{B-tree accessed bitmap join index.} \label{fig:bitmap_join}
%\end{figure}

The descent cost in the b-tree depends on its height. The b-tree's height built
on attribute $A$ is $log_{m}|A|$, where $m$ is the b-tree's order. This order
is equal to $K+1$, where $K$ represents the number of search keys in each
b-tree node. $K$ is equal to $\frac{S_{p}}{w(A)+S_{pointer}}$, where $w(A)$ and
$S_{pointer}$ are respectively the size of the indexed attribute $A$ and the
size of a disk page pointer in bytes. Without adding the b-tree leaf node
level, the b-tree descent cost is then $C_{descent}=log_{m}|A|-1$.

The scanning cost of leaf nodes is $\frac{|A|}{m-1}$ (at worst, all leaf nodes
are read). Data access is achieved through bits set to one in each bitmap. In
this case, it is necessary to read each bitmap. The reading cost of $d$ bitmaps
is $d\frac{|F|}{8S_{p}}$. Hence, the scanning cost of the leaf nodes is
$C_{scan}=\frac{|A|}{m-1}+d\frac{|F|}{8S_{p}}$.

The reading cost of the indexed table's tuples is computed as follow. For a
bitmap index built on attribute $A$, the number of read tuples is equal to
$\frac{|F|}{|A|}$ (if data are uniformly distributed). Generally, the total
number of read tuples for a query using $d$ bitmaps is
$N_{r}=d\frac{|F|}{|A|}$. Knowing the number of read tuples, the number of I/Os
in the reading phase is $C_{read}=p_{F}(1-e^{-
\frac{N_{r}}{p_{F}}})$~\cite{nei97imp}, where $p_{F}$ denotes the number of
pages needed for store the fact table.

In summary, the evaluation cost of a query exploiting a bitmap join index is
$C_{index}=log_{m}|A|-1+\frac{|A|}{m-1}+d \frac{|F|}{8S_{p}}+p_{F}(1-e^{-
\frac{N_{r}}{p_{F}}})$.

\subsection{Join cost without indexes}

If the bitmap join indexes are not useful while evaluating a given query, we
assume that all joins are achieved by the hash--join method. The number of I/Os
needed for joining table $R$ with table $S$ is then
$C_{hash}~=~3~(p_{S}~+~p_{R})$~\cite{mis92joi}.

\section{Bitmap join index selection strategy}\label{Strategy}

Our index selection strategy proceeds in several steps. The candidate index set
is first built from the frequent itemsets mined from the workload
(Section~\ref{Index selection}). A greedy algorithm then exploits an objective
function based on our cost models (Section~\ref{Cost models}) to prune the
least advantageous indexes. The detail of these steps and the construction of
the objective function are provided in the following sections.

\subsection{Candidate index set construction} \label{Construction}

From the frequent itemsets (Section~\ref{Index selection}) and the data warehouse
schema (foreign keys of the fact table, primary keys of the dimensions, etc.),
we build a set of  candidate indexes.

The SQL statement for building a bitmap join index is composed of three
clauses: \textbf{\texttt{On}}, \textbf{\texttt{From}} and
\textbf{\texttt{Where}}. The \textbf{\texttt{On}} clause is composed of
attributes on which is built the index (non--key attributes in the dimensions),
the \textbf{\texttt{From}} clause contains all joined tables and the
\textbf{\texttt{Where}} clause contains the join predicates.
%Figure~\ref{fig:bi_sql} shows an example of the Oracle~$9i$ SQL statement that
%builds a bitmap join index on the attribute \textbf{\texttt{FISCAL\_YEAR}} from
%the dimension \textbf{\texttt{Times}} (\textbf{\texttt{Sales}} is the fact
%table).
%\begin{figure}
%\begin{center}
%\begin{tabular}{p{11cm}}
%{\tt CREATE BITMAP INDEX BIJ\_TIMES ON Sales (Times.Fiscal\_Year)
%\newline FROM Sales,Times WHERE Sales.Time\_ID=Times.Time\_ID} \\
%\end{tabular}
%\end{center}\caption{SQL statement for building a bitmap join index}\label{fig:bi_sql}
%\end{figure}

We consider a frequent itemset $<Table.attribute_{1},...,Table.attribute_{n}>$
composed of elements such as $Table.attribute$. Each itemset is analyzed to
determine the different clauses of the corresponding index. We first extract
the elements containing foreign keys of the fact table because they are
necessary to define the \textbf{\texttt{From}} and \textbf{\texttt{Where}} index
clauses. Next, we retrieve the itemset elements that contain primary keys of
dimensions to form the \textbf{\texttt{From}} index clause. The elements containing
non--key attributes of dimensions form the \textbf{\texttt{On}} index clause. If
such elements do not exist, the bitmap join index cannot be built. %For example, the frequent itemset
%\textbf{\texttt{<Times.Time\_id}},\textbf{\texttt{Sales.Time\_id}},
%\textbf{\texttt{Times.Fiscal\_Year>}} generates the bitmap join index shown in
%figure~\ref{fig:bi_sql}.

\subsection{Objective functions}

In this section, we describe three objective functions to evaluate the variation
 of query execution cost, in number of I/Os, induced by adding a
new index. The query execution cost is assimilated to computing the cost of
hash joins if no bitmap join index is used or to the data access cost through
indexes otherwise. The workload execution cost is obtained by adding all
execution costs for each query within this workload. The first objective
function advantages the indexes providing more profit while executing queries,
the second one advantages the indexes providing more benefit and occupying less
storage space, and the third one combines the first two in order to select at
first all indexes providing more profit and then keep only those occupying less
storage space when this resource becomes critical. The first function is useful
when storage space is not limited, the second one is useful when storage space
is small and the third one is interesting when this storage space is quite
large. The detail of computing each function is not given due to the lack of
space.
%Appendix~\ref{ap:objective}.

\subsection{Index configuration construction}

The index selection algorithm is based on a greedy search within the candidate
index set $I$ given as an input. The objective function $F$ must be one of the
functions:  profit ($P$), profit/space ratio ($R$) or hybrid ($H$). If $R$ is
used, we add to the algorithm's input the space storage $M$ allotted for
indexes. If $H$ is used, we also add threshold $\alpha$ as input.

In the first algorithm iteration, the values of the objective function are
computed for each index within $I$. The execution cost of all queries in
workload $Q$ is equal to the total cost of hash joins. The index $i_{max}$ that
maximizes $F$, if it exists, is then added to the set of selected indexes $S$.
If $R$ or $H$ is used, the whole space storage $M$ is decreased by the amount
of space occupied by $i_{max}$.

The function values of $F$ are then recomputed for each remaining index in
$I-S$ since they depend on the selected indexes present in $S$. This helps
taking into account the interactions that probably exist between the indexes.
We repeat these iterations until there is no improvement or all indexes have
been selected ($ I-S= \emptyset $). If functions $R$ or $H$ are used, the
algorithm also stops when storage space is full.

%\begin{algorithm}[t]
%\caption{\textit{Index construction algorithm}}\label{algo:sel}
%\begin{algorithmic}

%\STATE  $S \leftarrow \emptyset$ \REPEAT

%\STATE $i_{max} \leftarrow \emptyset$

%\STATE $F_{max} \leftarrow 0$

%\FORALL{$i_{j} \in I-S$}

%\IF {$F_{/S}(i_{j}) > F_{max}$}

%\STATE $F_{max} \leftarrow F_{/S}(i_{j})$

%\STATE $i_{max} \leftarrow i_{j}$

%\ENDIF

%\ENDFOR

%\IF{$F_{/S}(i_{max}) > 0$}

%\STATE $S \leftarrow  S \cup \{i_{max}\}$

%\ENDIF

%\UNTIL{($F_{/S}(i_{max}) \leq 0$ ou $I-S=\emptyset$)}
%\end{algorithmic}
%\end{algorithm}

\section{Experiments}\label{Experiments}

In order to validate our bitmap join index selection strategy, we have run
tests on a data warehouse implemented within Oracle $9i$, on a Pentium
2.4~GHz~PC with a 512~MB main memory and a 120~GB~IDE disk. This data warehouse
is composed of the fact table \textbf{\texttt{Sales}} and five dimensions
\texttt{\textbf{Customers}}, \textbf{\texttt{Products}},
\textbf{\texttt{Promotions}}, \textbf{\texttt{Times}} and
\textbf{\texttt{Channels}}. We have measured for different value of the minimal
support parameterized in Close the workload execution time. In practice, the
minimal support limits the number of candidate indexes to generate and selects
only those that are frequently used.

For computing the different costs from our models, we fixed the value of
$S_{p}$ (disk page size) and $S_{pointer}$ (page pointer size) to 8 MB and 4 MB
respectively. These values are those indicated in the Oracle $9i$ configuration
file. The workload is composed of forty decision--support queries containing
several joins. We measured the total execution time when building indexes or
not. In the case of building indexes, we also measured the total execution time
when we applied each objective function among of  profit, ratio profit/space
and hybrid. We also measured the disk space occupied by the selected indexes.
When applying the cost models, we reduce the number of indexes and thereby the
storage space needed to store these indexes.

\begin{figure}[t]
 \begin{minipage}[b]{.5\linewidth}
  \centering\epsfig{figure=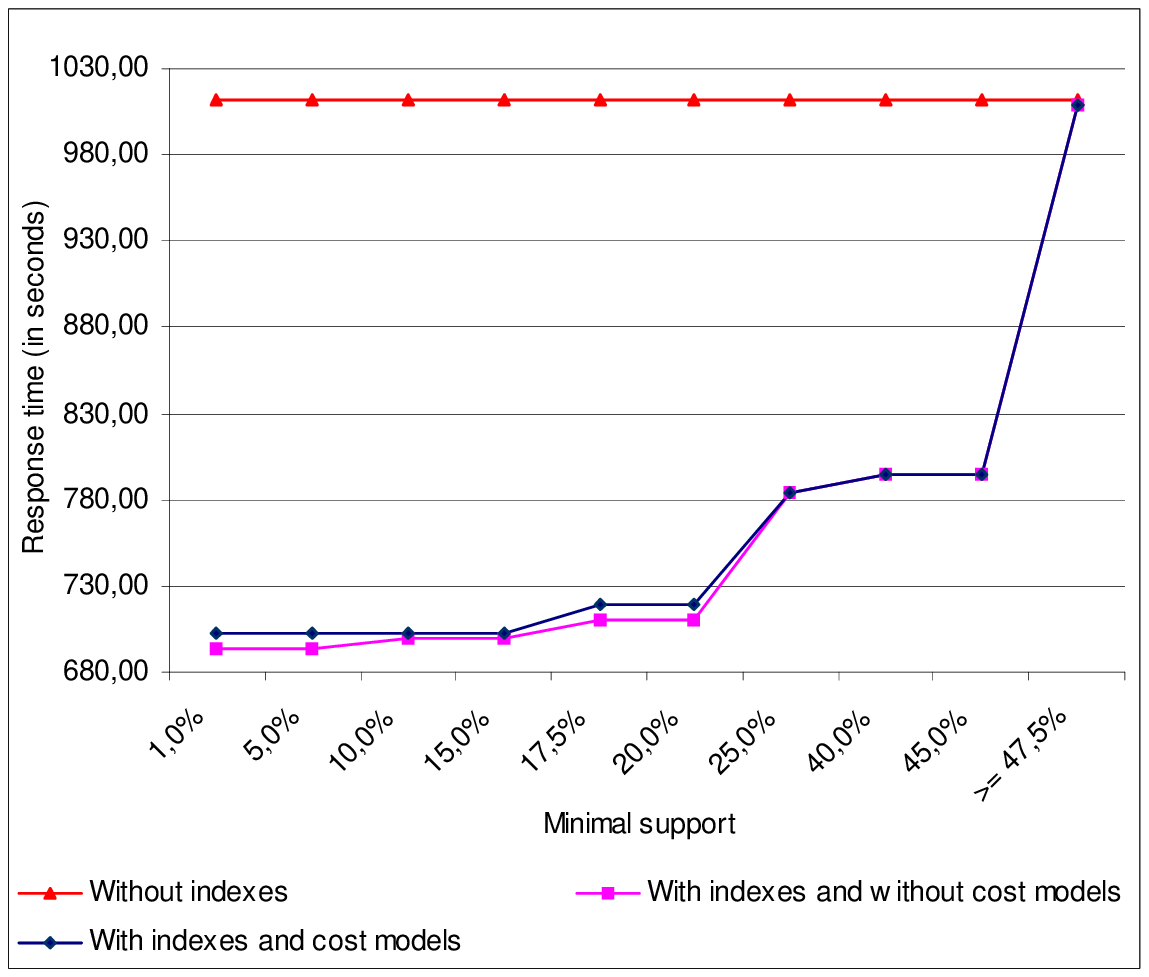,width=\linewidth}
  \caption{Profit function \label{fig:exp1}}
 \end{minipage} \hfill
 \begin{minipage}[b]{.5\linewidth}
  \centering\epsfig{figure=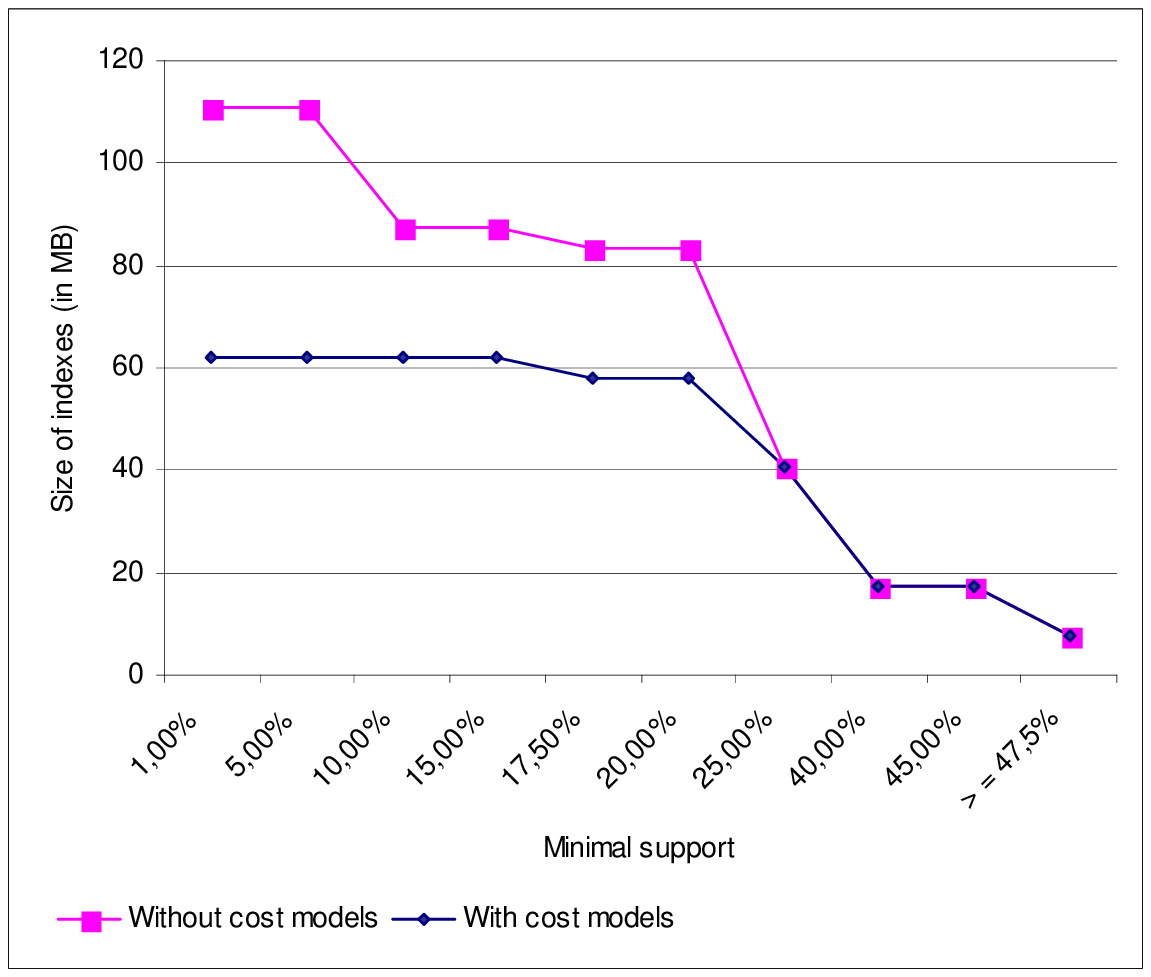,width=\linewidth}
  \caption{Index storage space \label{fig:exp2}}
 \end{minipage}
\end{figure}

\noindent \textbf{Profit function experiment.} Figure~\ref{fig:exp1} shows that
the selected indexes improve query execution time with and without application
of our cost models until the minimal support forming frequent itemsets reaches
47.5\%. Moreover, the execution time decreases continuously when the minimal
support increases because the number of indexes decreases. For high values of
the minimal support (greater than 47.5\%), the execution time is closer to the
one obtained without indexes. This case is predictable because there is no or
few candidate indexes to create. The maximal gain in time in both cases is
respectively 30.50\% and 31.85\%. Despite of this light drop of 1.35\% in time
gain when the cost models are used (fewer indexes are built), we observe a
significant gain in storage space (equal to 32.79\% in the most favorable case)
as shown in figure~\ref{fig:exp2}. This drop in number of indexes is
interesting when the data warehouse update frequency is high because update
time is proportional to the number of indexes. On the other hand, the gain in
storage space helps limiting the storage space allotted for indexes by the
administrator.

\begin{figure}[t]
 \begin{minipage}[b]{.5\linewidth}
  \centering\epsfig{figure=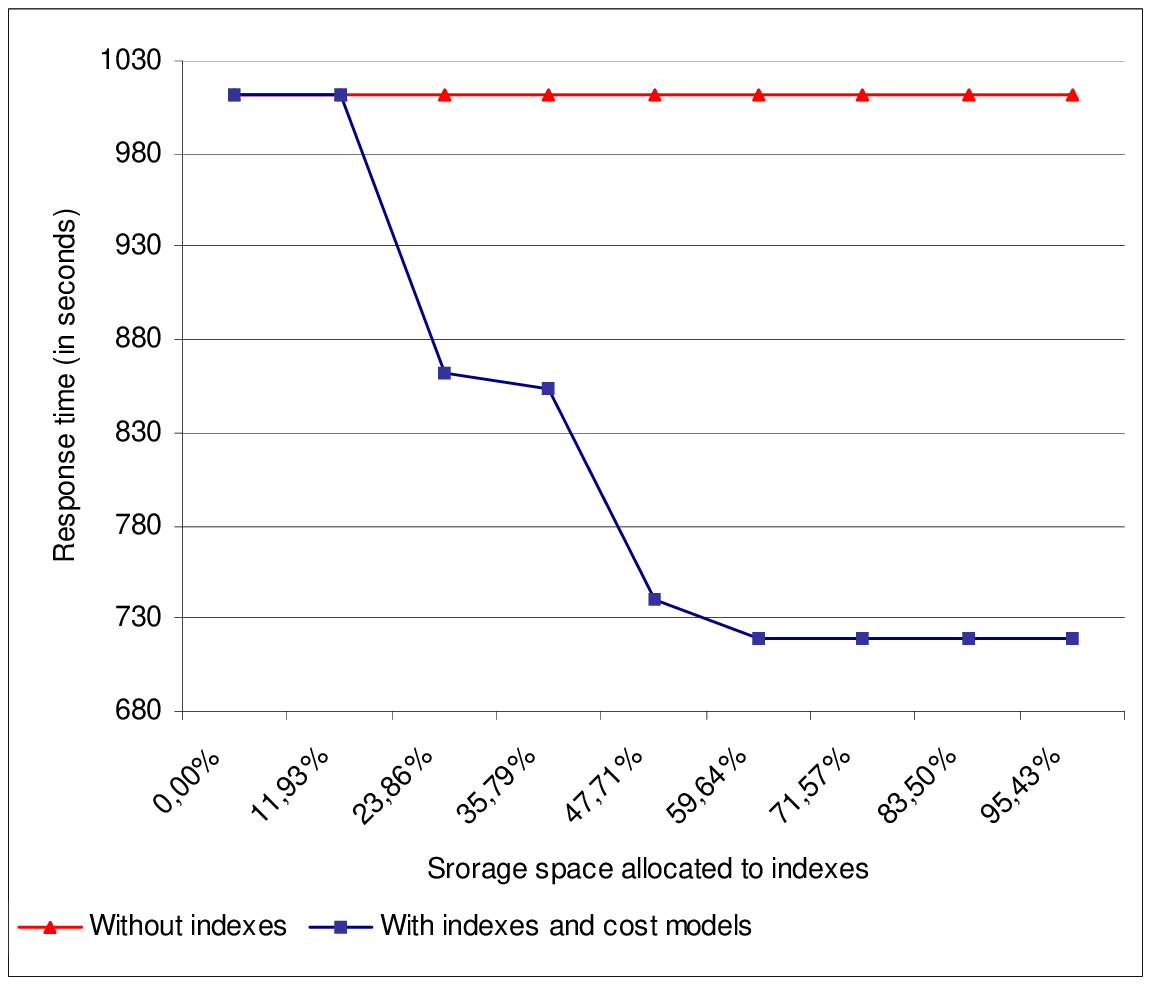,width=\linewidth}
  \caption{Profit/space ratio function \label{fig:exp3}}
 \end{minipage} \hfill
 \begin{minipage}[b]{.5\linewidth}
  \centering\epsfig{figure=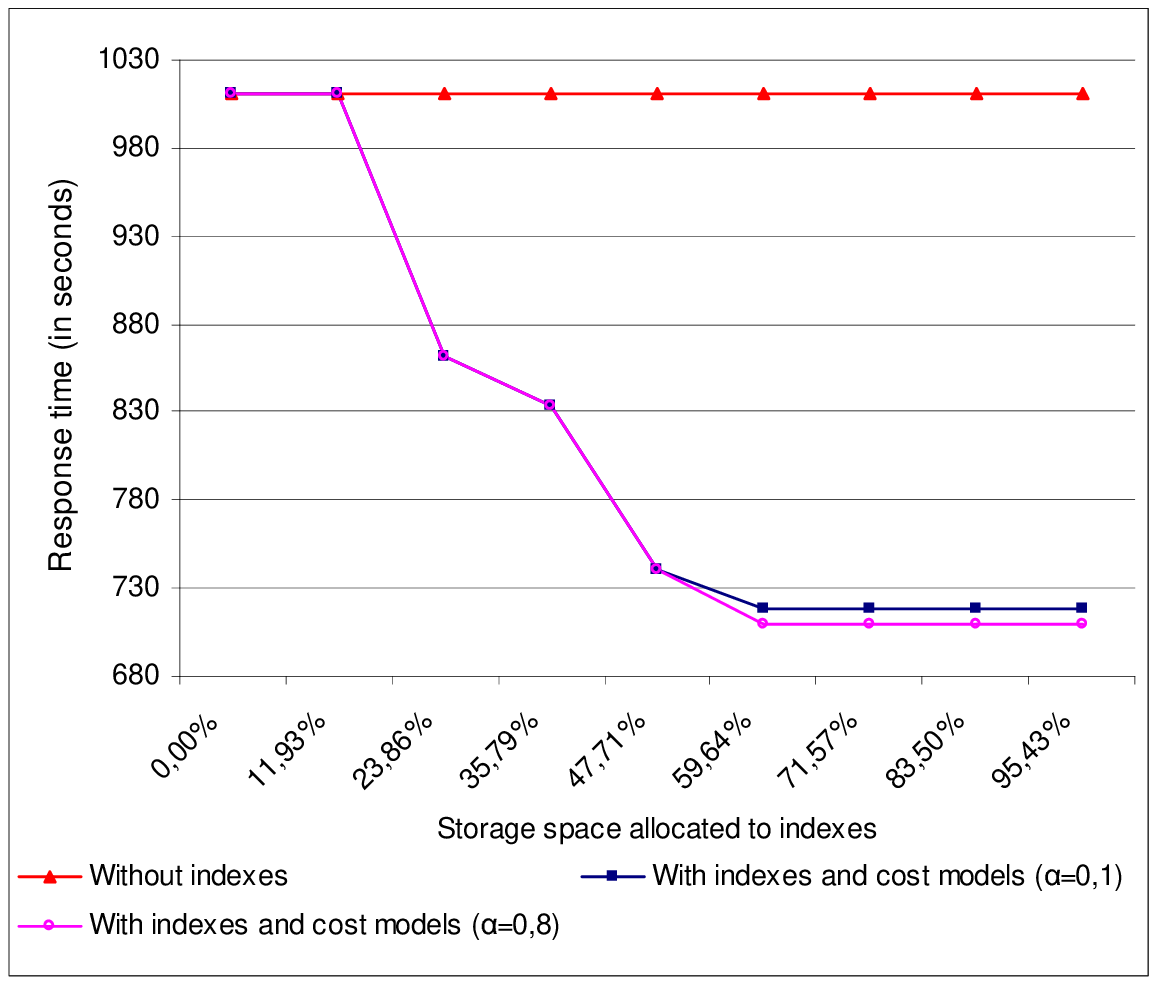,width=\linewidth}
  \caption{Hybrid function \label{fig:exp4}}
 \end{minipage}
\end{figure}

\noindent \textbf{Profit/space ratio function experiment.} In these
experiments, we have fixed the value of minimal support to 1\%. This value
gives the highest number of frequent itemsets and consequently the highest
number of candidate indexes. This helps varying storage space within a wider
interval. We have measured query execution time according to the percentage of
storage space allotted for indexes. This percentage is computed from the space
occupied by all indexes. Figure~\ref{fig:exp3} shows that execution time
decreases when storage space occupation increases. This is predictable because
we create more indexes and thus better improve the execution time. We also
observe that the maximal time gain is equal to 28.95\% and it is reached for a
space occupation of 59.64\%. This indicates that if we fix space storage to
this value, we obtain a time gain close to the one obtained with the profit
objective function (30.50\%). This case is interesting when the administrator
does not have enough space to store all the indexes.

\noindent \textbf{Hybrid function experiment.} We repeated the previous
experiments with the hybrid objective function. We varied the value of
parameter $\alpha$ between 0.1 and 1 by 0.1 steps. The obtained results with
$\alpha \in [0.1,0.7]$ and $\alpha \in [0.8,1]$ are respectively equal to those
obtained with $\alpha = 0.1$ and $\alpha = 0.7$. Thus, we represent in
figure~\ref{fig:exp4} only the results obtained with $\alpha = 0.1$ and $\alpha
= 0.7$. This figure shows that for $\alpha = 0.1$, the results are close to
those obtained with profit/space ratio the function  ; and for $\alpha = 0.8$,
they are close to those obtained with the profit function. The maximal gain in
execution time is respectively equal to 28.95\% and 29.95\% for $\alpha=0.1$
and  $\alpha=0.8$. We explain these results by the fact that bitmap join
indexes built on several attributes need more storage space. However, as they
pre--compute more joins, they better improve the execution time. The space
storage allotted for indexes then fills up very quickly after a few iterations
of the greedy algorithm. This explains why the parameter $\alpha$ does not
significantly affect our algorithm and the experiment results.

\section{Conclusion and perspectives}\label{Conclusion}

In this article, we presented an automatic strategy for bitmap index selection
in data warehouses. This strategy first exploits frequent itemsets obtained by
the Close algorithm from a given workload to build a set of candidate bitmap
join indexes. With the help of cost models, we keep only the most advantageous
candidate indexes. These models estimate data access cost through indexes, as
well as maintenance and storage cost for these indexes. We have also proposed
three objective functions:  profit, profit/space ratio and hybrid that exploit
our cost models to evaluate the execution cost of all queries. These functions
are themselves exploited by a greedy algorithm that recommends a pertinent
configuration of indexes. This helps our strategy respecting constraints
imposed by the system (limited number of indexes per table) or the
administrator (storage space allotted for indexes). Our experimental results
show that the application of cost models to our index selection strategy
decreases the number of selected indexes without a significant loss in
performance. This decrease actually guarantees a substantial gain in storage
space, and thus a decrease in maintenance cost during data warehouse updates.

Our work shows that the idea of using data mining techniques for data warehouse
auto-administration is a promising approach. It opens several future research
axes. First, it is essential to keep on experimenting in order to better
evaluate system overhead in terms of index building and maintenance. It could
also be very interesting to compare our approach to other index selection
methods. Second, extending our approach to other performance optimization
techniques (materialized views, buffering, physical clustering, etc.) is
another promising perspective. Indeed, in a data warehouse environment, it is
principally in conjunction with other physical structures such as materialized
views that indexing techniques provide significant gains in performance. For
example, our context extraction may be useful to build clusters of queries that
maximize the similarity between queries within each cluster. Each cluster may
be then a starting point to materialize views. In addition, it could be
interesting to design methods to efficiently share the available storage space
between indexes and views.
%~\cite{bel00eff}.

%Finally, this work relates to classical data warehouses. However, the data
%exploited in decision--support processes are now more and more complex. We term
%complex data multi-format and/or multi-source and/or multi-structure and/or
%multi-modal and/or multi-version data. In a complex data warehouse context,
%problems related to auto-administration remain unsolved. Index selection
%techniques may be adapted to this new field.

%Bibliography
\bibliographystyle{abbrv}
\bibliography{dawak}

\begin{thebibliography}{10}

\bibitem{agr00aut}
S.~Agrawal, S.~Chaudhuri, and V.~Narasayya.
\newblock Automated selection of materialized views and indexes in {SQL}
  databases.
\newblock In {\em 26th International Conference on Very Large Data Bases {(VLDB
  2000)}, Cairo, Egypt}, pages 496--505, 2000.

\bibitem{agr01mat}
S.~Agrawal, S.~Chaudhuri, and V.~Narasayya.
\newblock Materialized view and index selection tool for {M}icrosoft {SQL}
  {S}erver 2000.
\newblock In {\em {ACM SIGMOD} International Conference on Management of Data
  ({SIGMOD} 2001), Santa Barbara, USA}, page 608, 2001.

\bibitem{aou03fre}
K.~Aouiche, J.~Darmont, and L.~Gruenwald.
\newblock Frequent itemsets mining for database auto-administration.
\newblock In {\em 7th International Database Engineering and Application
  Symposium {(IDEAS 2003)}, Hong Kong, China}, pages 98--103, 2003.

\bibitem{cha04ind}
S.~Chaudhuri, M.~Datar, and V.~Narasayya.
\newblock Index selection for databases: A hardness study and a principled
  heuristic solution.
\newblock {\em IEEE Transactions on Knowledge and Data Engineering},
  16(11):1313--1323, 2004.

\bibitem{fel03nea}
Y.~Feldman and J.~Reouven.
\newblock A knowledge--based approach for index selection in relational
  databases.
\newblock {\em Expert System with Applications}, 25(1):15--37, 2003.

\bibitem{fin88phy}
S.~Finkelstein, M.~Schkolnick, and P.~Tiberio.
\newblock Physical database design for relational databases.
\newblock {\em ACM Transactions on Database Systems}, 13(1):91--128, 1988.

\bibitem{fra92ada}
M.~Frank, E.~Omiecinski, and S.~Navathe.
\newblock Adaptive and automated index selection in {RDBMS}.
\newblock In {\em 3rd International Conference on Extending Database Technology
  {({EDBT} 1992)}, Vienna, Austria}, volume 580 of {\em Lecture Notes in
  Computer Science}, pages 277--292, 1992.

\bibitem{gol02ind}
M.~Golfarelli, S.~Rizzi, and E.~Saltarelli.
\newblock Index selection for data warehousing.
\newblock In {\em 4th International Workshop on Design and Management of Data
  Warehouses ({DMDW} 2002), Toronto, Canada}, pages 33--42, 2002.

\bibitem{gup97ind}
H.~Gupta, V.~Harinarayan, A.~Rajaraman, and J.~D. Ullman.
\newblock Index selection for {OLAP}.
\newblock In {\em 13th International Conference on Data Engineering ({ICDE}
  1997), Birmingham, U.K.}, pages 208--219, 1997.

\bibitem{inm02bui}
W.~Inmon.
\newblock {\em Building the Data Warehouse}.
\newblock John Wiley \& Sons, third edition, 2002.

\bibitem{kim02dat}
R.~Kimball and M.~Ross.
\newblock {\em The Data Warehouse Toolkit: The Complete Guide to Dimensional
  Modeling}.
\newblock John Wiley \& Sons, second edition, 2002.

\bibitem{kra03gen}
J.~Kratica, I.~Ljubi\'c, and D.~To\v{s}i\'c.
\newblock A genetic algorithm for the index selection problem.
\newblock In {\em Applications of Evolutionary Computing, Essex, England},
  volume 2611 of {\em LNCS}, pages 281--291, 2003.

\bibitem{lab97phy}
W.~Labio, D.~Quass, and B.~Adelberg.
\newblock Physical database design for data warehouses.
\newblock In {\em 13th International Conference on Data Engineering ({ICDE}
  1997), Birmingham, U.K.}, pages 277--288, 1997.

\bibitem{mis92joi}
P.~Mishra and M.~Eich.
\newblock Join processing in relational databases.
\newblock {\em {ACM} Computing Surveys}, 24(1):63--113, 1992.

\bibitem{nei97imp}
P.~O'Neil and D.~Quass.
\newblock Improved query performance with variant indexes.
\newblock In {\em ACM SIGMOD International Conference on Management of Data
  ({SIGMOD} 1997), Tucson, USA}, pages 38--49, 1997.

\bibitem{pas99dis}
N.~Pasquier, Y.~Bastide, R.~Taouil, and L.~Lakhal.
\newblock Discovering frequent closed itemsets for association rules.
\newblock In {\em 7th International Conference on Database Theory {(ICDT
  1999)}, Jerusalem, Israel}, volume 1540 of {\em LNCS}, pages 398--416, 1999.

\bibitem{sar97ind}
S.~Sarawagi.
\newblock Indexing {OLAP} data.
\newblock {\em Data Engineering Bulletin}, 20(1):36--43, 1997.

\bibitem{val00db2}
G.~Valentin, M.~Zuliani, D.~Zilio, G.~Lohman, and A.~Skelley.
\newblock {DB2} advisor: An optimizer smart enough to recommend its own
  indexes.
\newblock In {\em 16th International Conference on Data Engineering ({ICDE}
  2000), San Diego, {USA}}, pages 101--110, 2000.

\bibitem{wu99que}
M.~Wu.
\newblock Query optimization for selections using bitmaps.
\newblock In {\em ACM SIGMOD International Conference on Management of Data
  ({SIGMOD} 1999), Philadelphia, USA}, pages 227--238, 1999.

\bibitem{wu98enc}
M.~Wu and A.~Buchmann.
\newblock Encoded bitmap indexing for data warehouses.
\newblock In {\em 14th International Conference on Data Engineering ({ICDE}
  1998), Orlando, USA}, pages 220--230, 1998.

\end{thebibliography}

\end{document}